\documentclass[preprint]{aastex}
\usepackage{rotating}
\usepackage{epsfig}  
\usepackage{epstopdf} 
\usepackage{graphics}
\usepackage{listings}

\begin{document}


\title{Direct Detection of the L-Dwarf Donor in WZ Sagittae}

\author{Thomas E. Harrison$^{\rm 1}$}

\affil{Department of Astronomy, New Mexico State University, Box 30001, MSC 
4500, Las Cruces, NM 88003-8001}

\begin{abstract}
Analysis of a large set of phase-resolved $K$-band spectra of the cataclysmic
variable WZ Sge shows that the secondary star of this system appears to be
an L-dwarf. Previous $K$-band spectra of WZ Sge found that the CO overtone
bandheads were in emission. We show that absorption from the 
$^{\rm 12}$CO$_{\rm (2,0)}$ bandhead of the donor star creates a dip in the 
$^{\rm 12}$CO$_{\rm (2,0)}$ emission feature. Measuring the motion of this 
feature over the orbital period, we construct a radial velocity curve that 
gives a velocity amplitude of K$_{\rm abs}$ = 520 $\pm$ 35 km s$^{\rm -1}$, 
consistent with the previously published values for this parameter.  
\end{abstract}


\begin{flushleft}
$^{\rm 1}$ Visiting Observer, W. M. Keck Observatory, which is operated 
as a scientific partnership among the California Institute of Technology, the 
University of California, and the National Aeronautics and Space 
Administration. NOAO Prop. \#2004B-0093\\
\end{flushleft}

\section{Introduction}

Cataclysmic variables (CVs) are short period binaries where the white dwarf 
primary is accreting matter from a lower mass secondary star.  An angular
momentum loss process keeps the system in contact (c.f., King \& Kolb
1995), and the orbital period shortens over time as the secondary continues to 
lose mass.  The standard evolutionary paradigm for CVs proposes that as these 
systems evolve to shorter and shorter periods, they eventually reach a period 
minimum somewhere between 70 and 80 minutes (Kolb \& Baraffe 1999, Howell et 
al. 1997, Knigge et al. 2011, and references therein). At this orbital period, 
the donor star has too low of a mass to support core hydrogen burning. After 
this point, the secondary can no longer adjust to the mass loss on its thermal 
time scale, and will expand upon further mass loss. The orbital period now must
increase. This sudden change to longer orbital periods has been termed 
the ``period bounce.'' One of the early lists for period bouncer candidates
was compiled by Patterson (1998). Patterson investigated the mass ratio 
distribution versus orbital period for the shortest period CVs, and found four 
good candidates for being period bouncers; one of these is WZ Sge.

WZ Sge is short period CV that exhibits infrequent, but large amplitude dwarf 
novae outbursts (Patterson et al. 2002). Whereas other short period CVs with 
similar orbital periods have outbursts on the weeks 
to months timescale, WZ Sge only has outbursts on a timescale of decades.
The orbital period of WZ Sge is accurately known from bright spot eclipse  
events (P$_{\rm orb}$ = 0.056687846 d; Patterson et al. 1998). The presence 
of ``superhumps'' in its outburst light curve 
(Patterson et al. 2002) is taken as evidence for a small mass ratio, if the 
thermal-tidal instability theory of Osaki (1996) is correct. Assuming an 
appropriate mass for the white dwarf suggests a secondary star close to the 
stellar/sub-stellar boundary. Such an object should have prominent CO 
absorption features in its $K$-band spectra, but to date, such observations 
have revealed that WZ Sge instead exhibits {\it CO emission features} whose
origin appears to be the accretion disk (Howell et al. 2004). Thus, 
even though WZ Sge is the closest known CV at 43 pc (Thorstensen  2003, 
Harrison et al.  2004a) and has a very low accretion rate (Patterson 1998), 
the donor object has remained elusive.

While direct photospheric absorption features from the donor in WZ Sge
have not previously been detected, H$\alpha$ emission, due to the irradiation 
of the companion during the 2001 outburst, were clearly seen, and led to
the first quantitative constraints on the nature of the secondary. 
Steeghs et al. (2001) found that the narrow H$\alpha$ emission feature leads 
to a radial 
velocity semi-amplitude of K$_{\rm em}$ = 493 $\pm$ 10 km s$^{\rm -1}$ for the
donor. Modeling the same data set gives K$_{\rm 1}$ = 37 $\pm$ 5 km 
s$^{\rm -1}$. Incorporation of systematic errors leads to 493 km s$^{\rm -1}$ 
$\leq$ K$_{\rm em}$ $\leq$ 585 km s$^{\rm -1}$, and a mass ratio range of 0.04 
$\leq$ $q$ $\leq$ 0.075. With M$_{\rm 1}$ $>$ 0.70 M$_{\sun}$, the secondary 
has M$_{\rm 2}$ $<$ 0.11 M$_{\sun}$, consistent with an object near the 
stellar/substellar boundary (Dupuy et al. 2010).  
Harrison et al. (2013) modeled infrared light curves and optical/IR photometry 
of WZ Sge and found that both suggest a secondary with a 
temperature near T$_{\rm eff}$ = 1800 K. They found that an L2 dwarf (with
M$_{\rm K}$ = 11.1) that supplied $\sim$ 60 \% of the $K$-band flux was
consistent with the infrared light curves.

Given these results, the secondary star of WZ Sge {\it should} be visible in 
$K$-band spectra if it is of normal composition. However, with CO emission, 
and the possibility of sub-solar carbon abundances (see Harrison \& Hamilton 
2015), a more extensive set of data is required to search for the donor. We 
describe below a large set of $K$-band spectra that clearly reveal the L-dwarf 
secondary in WZ Sge. In the next section we describe the observations, in 
section 3 we detail our analysis and results, and draw our conclusions in 
section 4.

\section{Observations}

$K$-band spectroscopy of WZ Sge was obtained using NIRSPEC on $Keck$ II on
the night of 2004 August 28. We used NIRSPEC in single order, low resolution
mode with the
0.38" slit, resulting in a dispersion of 4.23 \AA/pix (R = 2300).
WZ Sge was observed continuously from 05:23 UT to 07:37 UT and from 08:20 UT 
until 11:30 UT. The 43 minute gap is due to the fact that WZ Sge transits the 
zenith as seen from Mauna Kea, and thus cannot be tracked by an alt-az 
telescope. We were on source for 5.4 hr, obtaining near complete coverage of 
four orbital periods.  Eighty eight spectra were obtained. The exposure times 
for all of the WZ Sge spectra were 180 s. The raw spectra have S/N $\sim$ 30 
in the brightest portions of the continuum. Before observing WZ Sge, during 
the transit, and immediately afterwards, arc lamps, flats and telluric 
standard star data were obtained. The observations were reduced in the normal 
way using IRAF. 
Bright G2V stars were observed to provide for telluric correction for the 
wide range of airmasses covered by our observations of WZ Sge (1.53 $\leq$
sec$z$ $\leq$ to 1.0). We used G2V telluric standards in an attempt to better 
recover the H I Br$\gamma$ profile, using the procedure outlined
in Maiolino et al. (1996): After division by the G2V telluric, the
spectrum is then multiplied by a rotationally convolved solar spectrum 
(see Harrison et al.  2004b). The data were phased to the ephemeris of 
Patterson et al. (1998), but with the offset in the timing of inferior 
conjunction found by Harrison et al. (2013).

\section{Results}

\begin{figure}[htb]
\centerline{{\includegraphics[width=15cm]{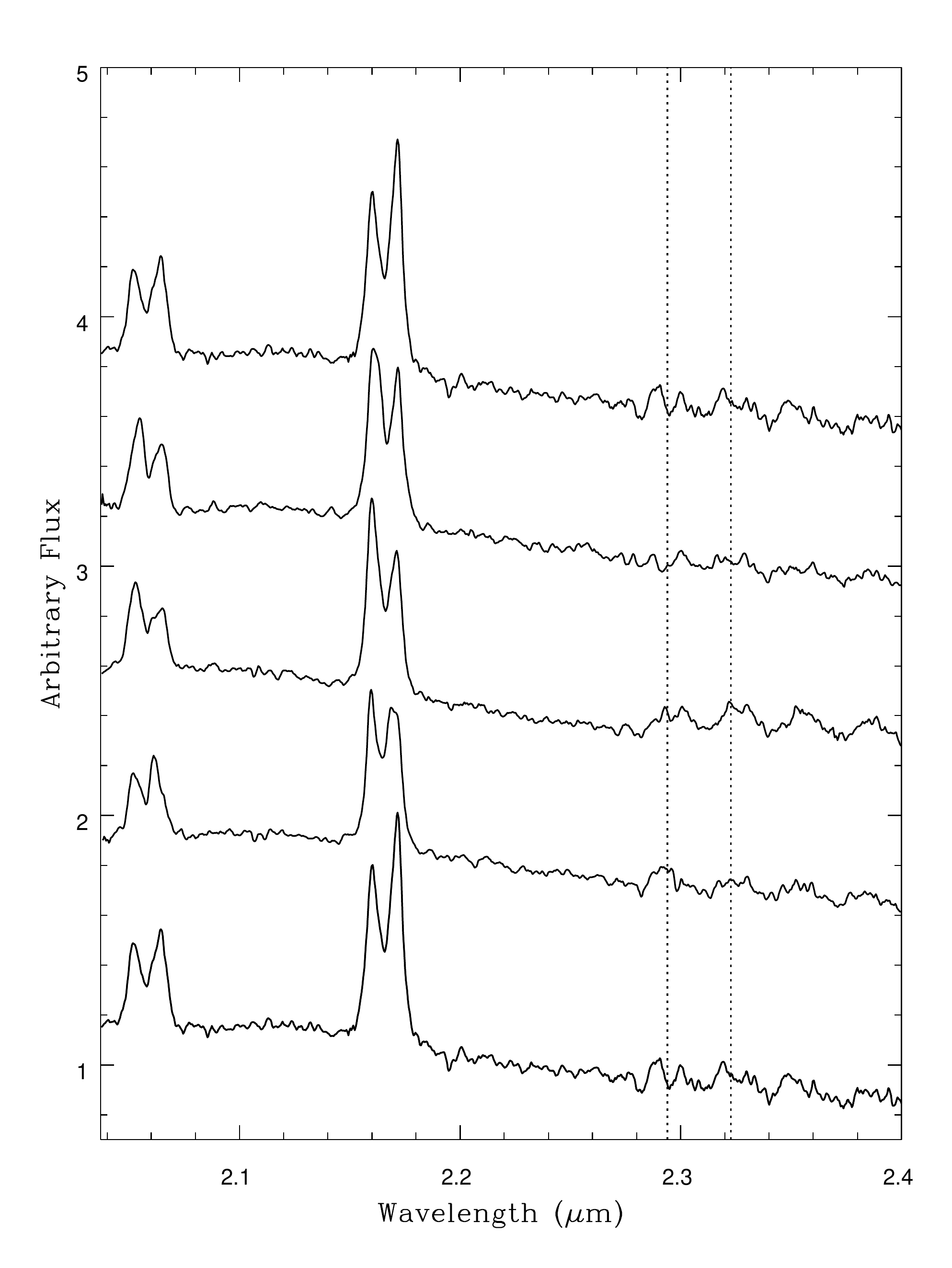}}}
\caption{The $K$-band spectra of WZ Sge at $\phi$ = 0.0 (bottom), 0.25,
0.5, 0.75 and 0.0 (top). The spectra presented in this figure have been 
boxcar smoothed by 5 pixels. The location of the $^{\rm 12}$CO$_{\rm (2,0)}$
($\lambda$2.294 $\mu$m) and $^{\rm 12}$CO$_{\rm (3,1)}$ ($\lambda$2.323 
$\mu$m) bandheads are delineated by vertical dotted lines.}
\label{4spec}
\end{figure}

In Fig. \ref{4spec} we present a subset of the data at the phases 0.0, 0.25, 
0.5 and 0.75 for comparison with the $K$-band spectra presented in Howell et 
al.  (2004). The CO emission features appear to be of the same strength for
both epochs, however, emission from H$_{\rm 2}$ is not present in the
new data. The main difference between the two data sets is the dramatic 
improvement in S/N afforded by having numerous redundant spectra to median
together. The phase 0.0 spectrum clearly shows an absorption dip centered in 
the middle of the first CO$_{\rm (2,0)}$ ($\lambda$2.294 $\mu$m) emission 
feature. At phase 0.25, the red side of the CO emission feature appears to 
dramatically decline as would be expected if the underlying absorption feature
of the secondary star was maximally red shifted at this time. The reverse 
happens at phase 0.75: the blue wing of the CO emission feature is now greatly 
weakened.  The changing morphology of the CO emission feature in Fig. 
\ref{4spec} suggests the presence of a late-type secondary. 

\subsection{A Radial Velocity Curve for the Donor}
\begin{figure}[htb]
\centerline{{\includegraphics[width=15cm]{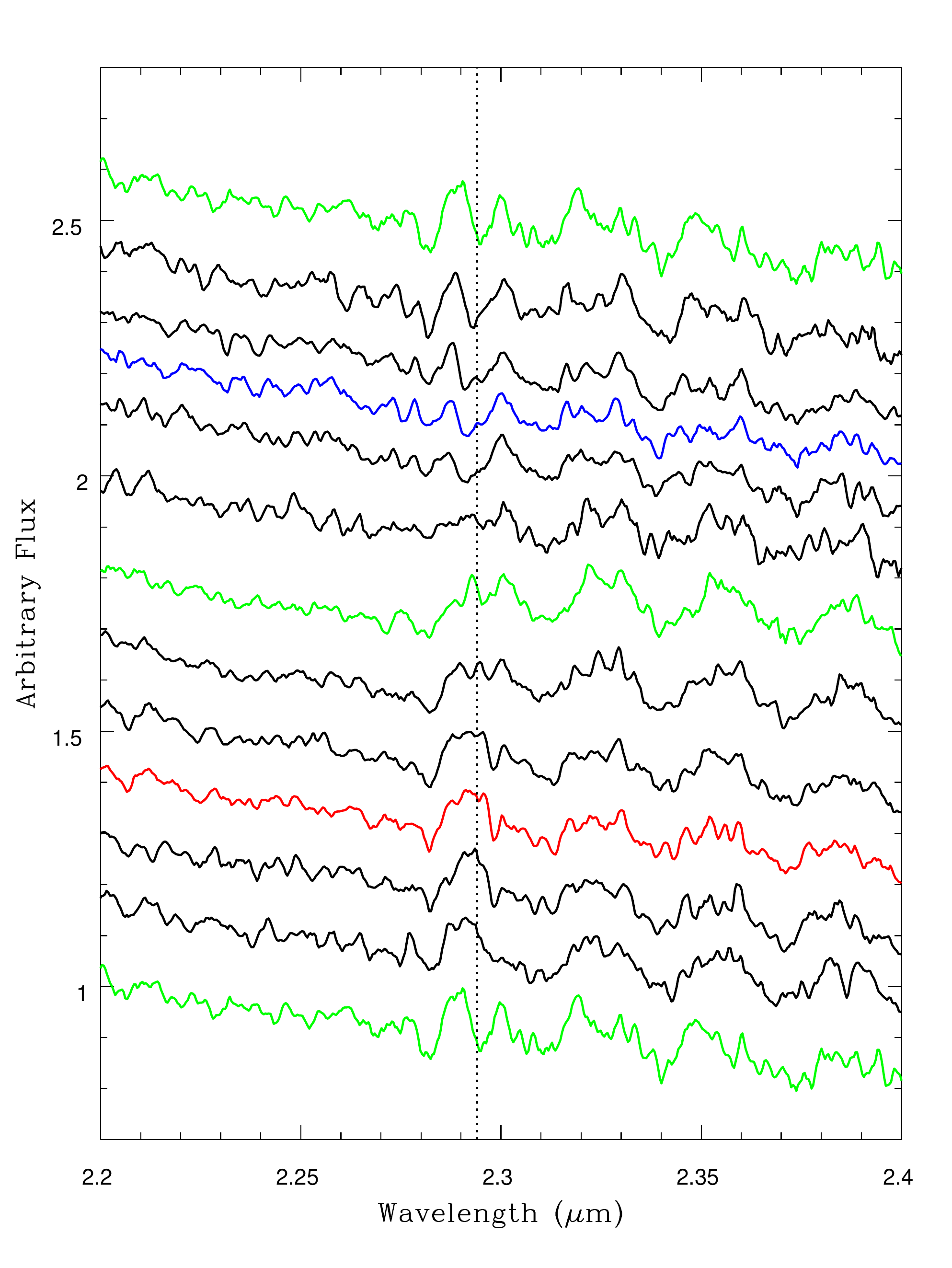}}}
\caption{The spectra of WZ Sge in the wavelength interval 2.2 $\leq$ $\lambda$
$\leq$ 2.4 $\mu$m. Phase 0 is at the bottom (green), and the spectra step by
$\Delta \phi$ = 0.1, except with the $\phi$ = 0.25 (red) and 0.75 (blue)
spectra inserted into the sequence. The $\phi$ = 0.5 spectrum is also 
plotted in green. The spectra presented in this figure have been boxcar 
smoothed by 5 pixels. The vertical dotted line delineates the location of
the $^{\rm 12}$CO$_{(2,0)}$ bandhead at 2.294 $\mu$m.  }
\label{coreg}
\end{figure}

To more closely examine the CO
region, we present Fig. \ref{coreg}. The spectra shown in this figure
are the medians of 7 to 11 individual spectra with $\Delta \phi$ = $\pm$ 0.05
in phase. This figure has data at every 0.1 in phase, with the addition
of the $\phi$ = 0.25 (red) and 0.75 (blue) spectra shown in Fig. \ref{4spec}.
The bandhead location for CO$_{\rm (2,0)}$ is delineated by the dotted
line. As is clear, the absorption dip is centered in the emission feature
at phase 0, moves redward to a maximum at phase 0.25, returns to no offset
at phase 0.5, and exhibits the maximum blueshift at phase 0.75.  

\begin{figure}[htb]
\centerline{{\includegraphics[width=16cm]{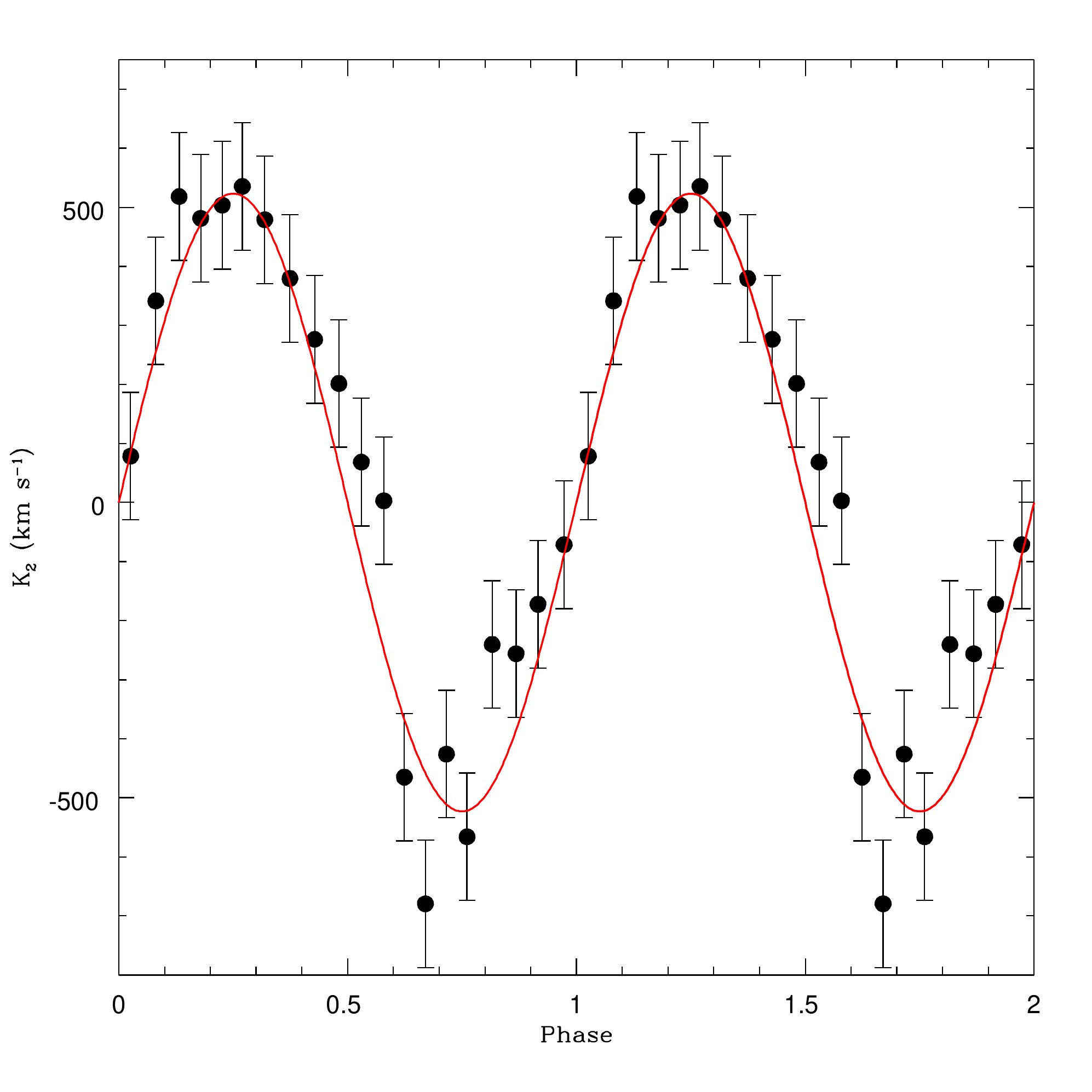}}}
\caption{The radial velocity measurements of the donor star derived from
the positions of the $^{\rm 12}$CO$_{(2,0)}$ bandhead. A sinusoid with
an amplitude of 520 km s$^{\rm -1}$ is plotted in red.}
\label{rv}
\end{figure}

\begin{figure}[htb]
\centerline{{\includegraphics[width=16cm]{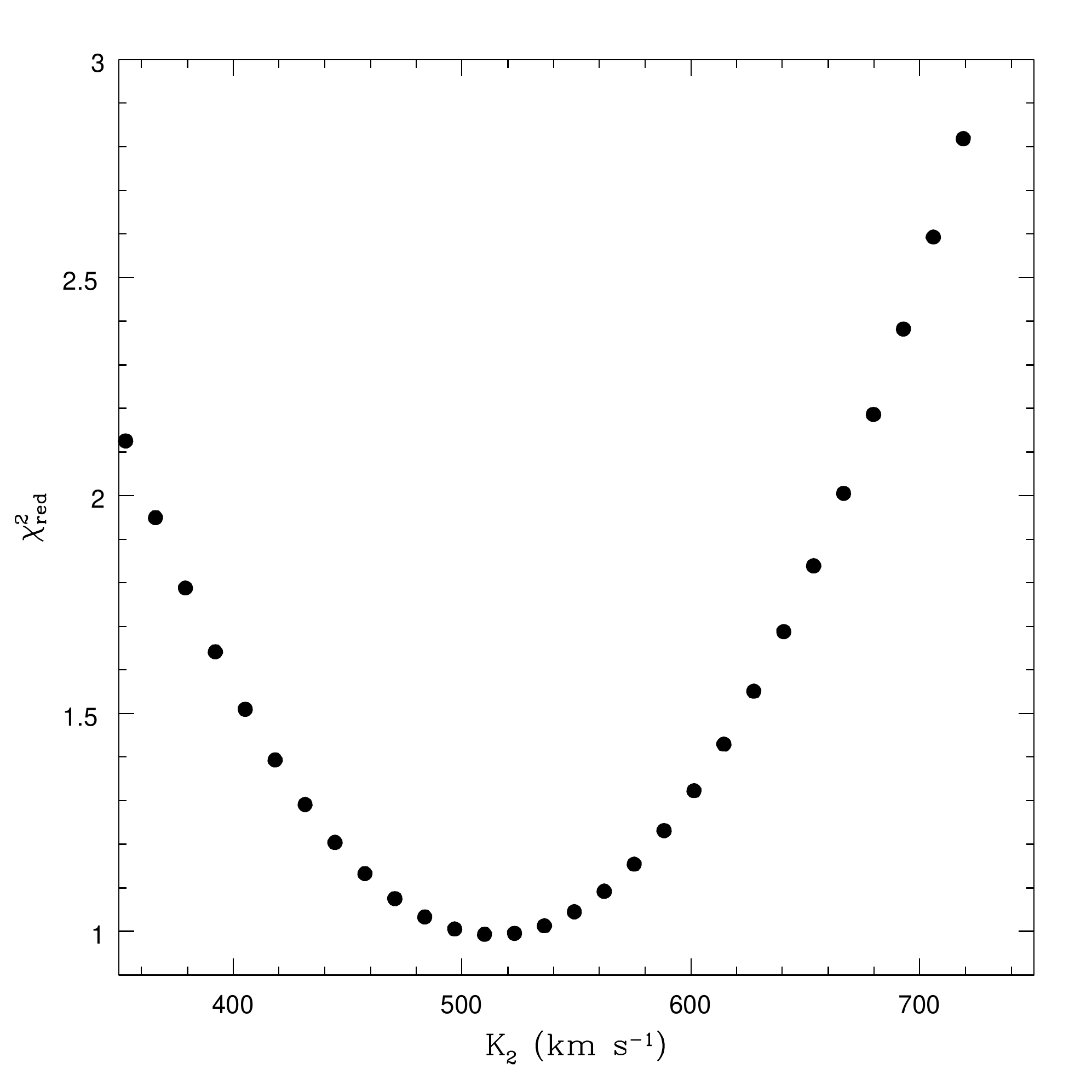}}}
\caption{A $\chi^{\rm 2}$ analysis of the radial velocity data finds that
the best fitting K$_{\rm 2}$ radial velocity semi-amplitude is near 520 km 
s$^{\rm -1}$.}
\label{chisq}
\end{figure}

To obtain a radial velocity curve, we rebin the data in $\Delta \phi$ = $\pm$ 
0.025 increments to reduce orbital smearing. We also average the phases of
the individual spectra to get a weighted mean value for the phase of each 
medianed data set. We used the line measurement capability in IRAF (``splot'') 
to fit a Gaussian profile to the observed absorption feature. The radial 
velocity curve we derive, shown in Fig. \ref{rv}, consists of 20 individual 
measurements. To arrive at this result, we subtracted the measured wavelength
of the CO bandhead in the individual spectra from its expected position
(2.294 $\mu$m). We then offset the data points by the mean of the entire radial
velocity data set. In these lower S/N spectra, the location of the first 
overtone CO bandhead is often poorly defined, leading to significant deviations 
from a smooth radial velocity curve. The origin of the most deviant 
measurements can be clearly traced back to the data shown in Fig. \ref{coreg} 
(such as near $\phi$ = 0.6). With the expectation of a sinewave, we performed 
a $\chi^{\rm 2}$ analysis of the data presented in Fig. \ref{rv}. The result, 
shown in Fig.  \ref{chisq}, finds a best fit velocity of K$_{\rm abs}$ = 520 
$\pm$ 35 km s$^{\rm -1}$. This value for K$_{\rm abs}$ is 
consistent with the range of velocities derived by Steeghs et al. (2001). A 
radial velocity curve with this amplitude is plotted in red in Fig. \ref{rv}. 
The $\chi^{\rm 2}$ fitting also allows us to assign error bars to our 
measurements, as to get values of $\chi^{\rm 2}_{\rm red}$ $\sim$ 1, 
requires $\sigma$ = 108 km s$^{\rm -1}$. This value for the error on the 
individual radial velocity measurements appears to represent the observed 
scatter quite well.

The exact value of the semi-amplitude of the secondary is difficult to 
constrain. Our exposures were three minutes in duration ($\Delta \phi$ = 0.04),
and our medianed spectra spanned $\Delta \phi$ $\leq$ 0.05. Thus, the short 
period of WZ Sge introduces orbital smearing into our results. Combined with 
our large error bars, the derived value of K$_{\rm abs}$ is not especially
precise. Careful examination of Fig. \ref{rv} reveals that the absorption 
features in the continuum between H I Br$\gamma$, and the first overtone 
bandhead of CO, also track the motion of the secondary. These features closely 
resemble those of an L2 dwarf. If we assume that these are from the donor
star, we can endeavour to determine an alternative value for K$_{\rm abs}$.
We attempted to use a cross-correlation technique using the L2 dwarf Kelu 1
(obtained from the IRTF Spectral Library\footnote{http://irtfweb.ifa.hawaii.edu/$\sim$spex/IRTF\_Spectral\_Library/}; Cushing 2005), as our template. 
Unfortunately, that effort was unsuccessful. We next decided to Doppler-correct 
the entire data set until these continuum features were at their sharpest. We 
performed 
a $\chi^{\rm 2}$ test over the velocity interval 490 $\leq$ K$_{\rm 2}$ $\leq$ 
575 km s$^{\rm -1}$, and over the wavelength range 2.19 $\leq$ $\lambda$ $\leq$ 
2.28 $\mu$m. The results, however, were ambiguous, with nearly identical 
values of $\chi^{\rm 2}$ throughout the range. The strongest minimum was at 
K$_{\rm abs}$ = 510 km s$^{\rm -1}$, with a secondary minimum at K$_{\rm abs}$ 
= 550 km s$^{\rm -1}$. A visual inspection suggests that 
K$_{\rm 2}$ = 555 km s$^{\rm -1}$ produces a spectrum whose features best match those of Kelu 1.

\begin{figure}[htb]
\centerline{{\includegraphics[width=17cm]{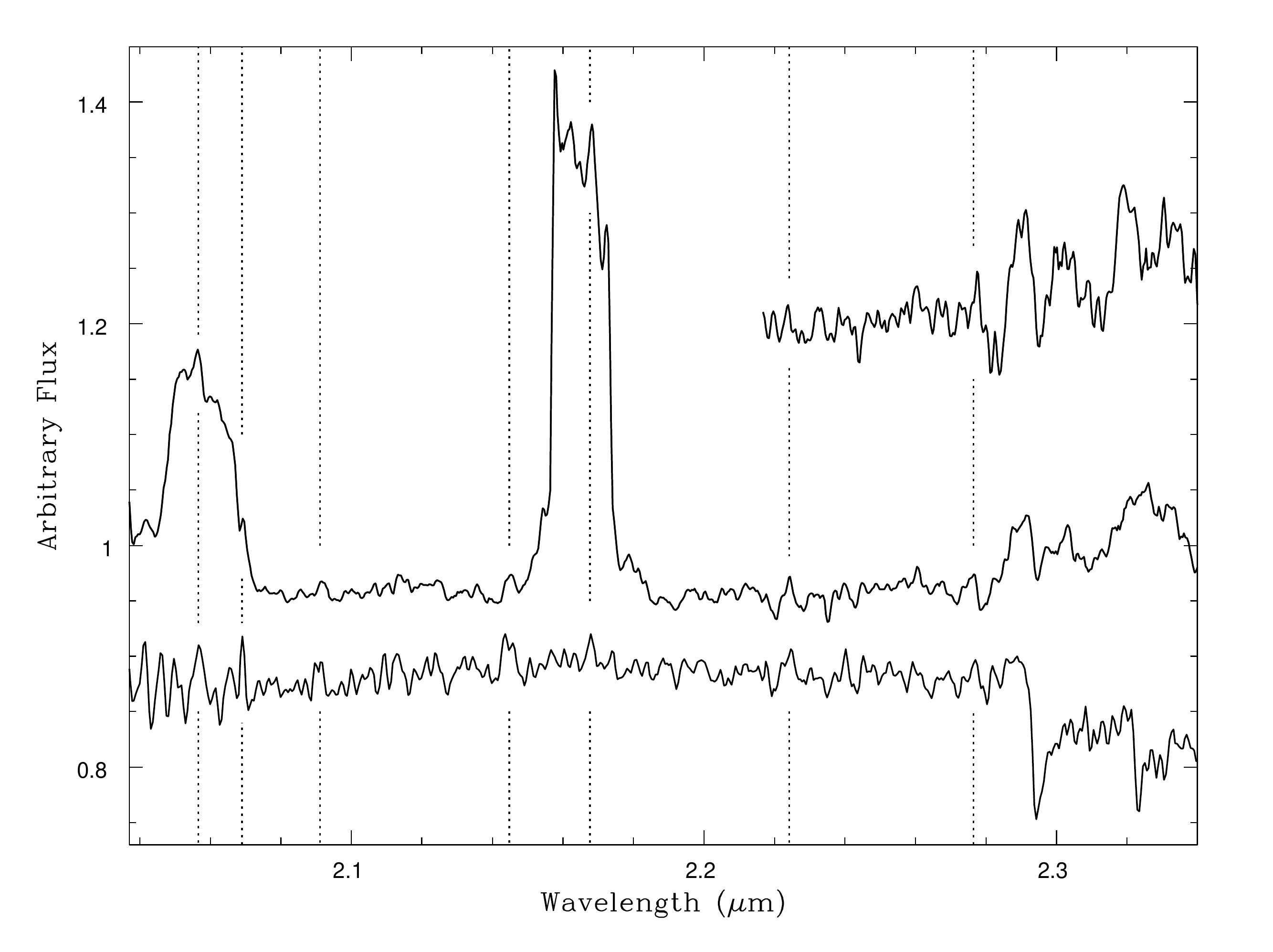}}}
\caption{The median of the entire data set for WZ Sge Doppler-corrected to 
K$_{\rm 2}$ = 555 km s$^{\rm -1}$, compared to the spectrum of the L2 dwarf 
Kelu 1. The vertical dotted lines locate opacity minima peaks in the L2 dwarf 
that are obviously present in the spectrum of WZ Sge.  Note that
both have been continuum subtracted, and the L2 dwarf spectrum then divided by
a factor of two. On the top-right we plot a segment of the median of the
spectra at phase 0.0 ($\Delta \phi$ = $\pm$0.05). The CO features are much
stronger near phase 0.0 due to the lack of irradiation, and the lower amount of
dilution due to the partial eclipse of the white dwarf and accretion disk
at this time.}
\label{ldwarf}
\end{figure}

In Fig. \ref{ldwarf}, we compare the Doppler-corrected data set (assuming
K$_{\rm 2}$ = 555 km s$^{\rm -1}$) to Kelu 1. Kelu 1 is in fact a 
binary consisting of $\sim$L2 and $\sim$L3 dwarfs (Liu \& Leggett 2005). The 
primary is about 0.5 mag brighter in the $K$-band than the secondary. Kelu 
1 is a rapid rotator, with a period of P$_{\rm rot}$ = 1.8 hr (Clarke et al.
2002), and $v$sin$i$ = 60 $\pm$ 5 km s$^{\rm -1}$ (Basri et al. 2000). With
K$_{\rm 2}$ = 555 km s$^{\rm -1}$, and the mean parameters for WZ Sge from
Steeghs et al. (2001), the rotation velocity of the secondary should be of order
$v$sin$i$ $\sim$ 85 km s$^{\rm -1}$. Neither rotational velocity is resolved
in these spectra, explaining their similarity in appearance. To construct Fig. 
\ref{ldwarf} we continuum fit and divided the spectra of both objects. In 
addition, we then divided the spectrum of Kelu 1 by a factor of 2.0 to best 
match the amplitude of the corresponding features in WZ Sge. This amount of 
``dilution'' is consistent with that found in the light curve modeling of WZ 
Sge by Harrison et al.  (2013). 

\subsection{The CO Features of the L Dwarf Donor}

In the mean of the full data set, the first overtone CO feature in WZ Sge 
certainly appears to be much weaker than that seen in Kelu 1. As noted in 
Harrison et al. (2013), light curve modeling indicates irradiation of the 
donor star near $\phi$ = 0.5. In Fig. \ref{ldwarf} we also plot a portion of 
the Doppler-corrected data set centered on $\phi$ = 0.0. The CO features
are considerably stronger at inferior conjunction perhaps partly due to the 
lack of irradiation at this phase, but mostly due to the dramatic decrease in 
the dilution from the partial eclipse of the accretion
disk and hotspot. The light curve modeling in Harrison et al. (2013) can 
provide insight on the level of irradiation. In their ellipsoidal
models, the secondary minimum ($\phi$ = 0.5) in the model $K$-band light curve
(their Fig. 12) is 0.2 mag less deep than it would be if there was no 
irradiation. If there is no irradiation, and the secondary dominated
the $K$-band luminosity, the secondary minimum would be slightly deeper
than the primary minimum (c.f., Fig. 14 in Harrison \& Campbell 2015). Thus, 
if the secondary minimum is weaker than the
primary minimum, it is an indication of irradiation. The change in temperature 
required to create the observed 0.2 mag difference, however, is very small 
($\sim$5\%). This is insufficient to significantly change the spectral type
of the donor and thus alter the strength of the CO feature.

We believe that there are two plausible scenarios that might explain why the CO 
absorption feature in the donor is especially weak near $\phi$ = 0.5. The 
first is that the CO emission appears to be strongest at this phase. If one 
compares the entire set of CO emission features they are all strongest at this 
phase. Why does this happen? It could be that irradiation from the hotspot is
suppressing CO emission from nearby regions in the accretion disk. Since the 
hotspot leads the secondary by $\Delta \phi \sim$0.05, if there is a decrease 
of CO emission near the hotspot, the red wing of the CO emission features
on the receding side of the disk near the hotspot would be diminished, and 
the blue wing would be lessened on the approaching part of its orbit. Near
phase 0.5, there would be no significant affect, and the profiles would
be at their strongest, and most symmetric. The morphology of the CO emission 
features shown in Fig. 2 are certainly consistent with this scenario. Note 
also that the center of the first overtone emission feature suggests that it 
is slightly redshifted at phase 0.5, and that the red wing of the same feature 
is weaker at phase 0. The small offset in phase of the hotspot certainly 
could explain the changing morphologies of the emission features. This phasing
offset 
would also produce an asymmetric irradiation of the secondary star. There might 
be evidence for this in the infrared light curves of WZ Sge. There is a 
weak excess in the $K$-band and $Spitzer$ IRAC 4.5 $\mu$m bandpass light 
curves of WZ Sge at $\phi$ = 0.7 (vs. 0.2), suggesting a tiny 
bit of additional irradiation of the hemisphere facing the observer at this 
time. This amount of irradiation appears to be much too small, however, to 
impart a significant change on the spectral properties of the secondary. 

Secondly, it might be possible for the hotspot to obscure the secondary
near superior conjunction. As noted by a number of authors (see Mason et al. 
2000 and references therein), the hotspot in WZ Sge has considerable optical 
depth. If we assume this remains true in the $K$-band, and that the hotspot has 
sufficient vertical extent, it might be able to obscure a small portion of the 
secondary star near $\phi$ = 0.5. The $K$-band light curve in Harrison et al. 
(2013) does show a sharper than expected secondary minimum. The presence
of strong CO emission and a small amount of obscuration could
combine to create the weaker than expected CO absorption features near
superior conjunction. 

\begin{figure}[htb]
\centerline{{\includegraphics[width=15cm]{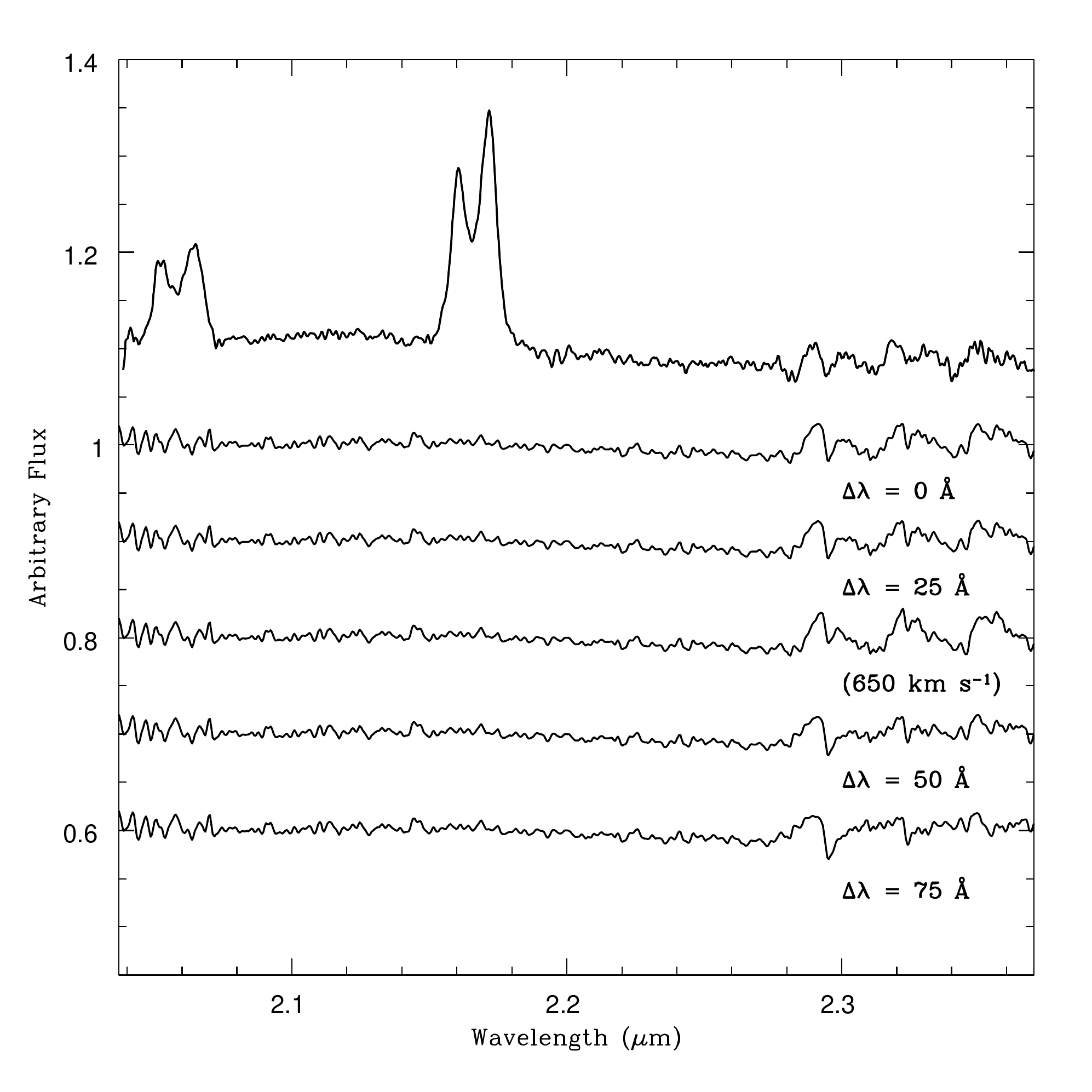}}}
\caption{Models where we have added a CO emission line spectrum to the L2 
dwarf data. The first model spectrum plotted below that of WZ Sge is a single 
CO emission line spectrum broadened by a Gaussian with a velocity of 1300 km 
s$^{\rm -1}$ (``$\Delta \lambda$ = 0 \AA'') added to the L2 dwarf spectrum. 
Clearly, this spectrum 
closely matches the observations and suggests a normal CO feature in the donor 
star of WZ Sge. Below that spectrum is plotted the addition of two CO emission
line spectra that have had their peaks shifted by $\pm$ 25 \AA, and then added 
to the spectrum of Kelu 1. At the 1300 km s$^{\rm -1}$ velocity broadening 
used here, there is very little difference between this model, and the single 
CO emission line model. If we cut the velocity broadening in half (650 km 
s$^{\rm -1}$), but keep the $\pm$ 25 \AA ~separation between the two CO 
spectra, we get a model that does a much poorer job at matching the 
observations.  Models with larger separations of the CO emission peaks 
cannot reproduce the observations.  }
\label{cocomp}
\end{figure}

We can attempt to model the depth of the CO absorption feature seen in the 
medianed full data set by adding a CO emission spectrum to the L2 data 
following the procedure used in Harrison \& Hamilton (2015, see their Fig. 21). 
In the case of WZ Sge, there is the expectation that the CO features 
will have a double-peaked profile similar to the H I and He I lines.  
Since it is impossible to know the exact splitting a priori, we model
several cases. The two peaks of the H I lines (see Fig. 1) are separated by 
105 \AA. If the
CO emission comes from an outer, cooler region of the disk, the splitting
should be less than this. A single Gaussian fit to the CO$_{\rm (2,0)}$ feature
leads to an estimate of the velocity broadening of $\sim$ 1300 km s$^{\rm -1}$. 
In Fig. \ref{cocomp}, we have added a CO emission line spectrum with this 
amount of velocity broadening to the IRTF spectrum of Kelu 1. Next we add
two velocity-broadened CO emission spectra shifted by $\pm$ 25 \AA 
~(total $\Delta \lambda = 50$ \AA) to the L2 dwarf spectrum, and plot it just 
below the single CO emission line spectrum. 
Below that is a CO emission spectrum also shifted
by $\pm$ 25 \AA , but with each emission component broadened by 650 km 
s$^{\rm -1}$. As shown in Harrison \& Hamilton (2015), as the Gaussian profile 
of the CO emission narrows, the observed depth of the CO feature decreases due 
to the emission feature contributing more flux into the deepest parts of the 
absorption bandhead of the secondary star. Note that at this velocity, the 
blue wing does not match the observations. Below this spectrum in Fig. 
\ref{cocomp} we plot the sum of a CO emission spectrum and the L2 dwarf
where the separation between the two peaks is $\Delta \lambda$ = $\pm$ 50 \AA, 
and below that, one with $\pm$ 75 \AA ~separation. At a velocity broadening of 
1300 km s$^{\rm -1}$, splittings of $\leq$ $\pm$ 50 \AA ~ are too small to 
actually affect the CO emission profile. As we move to larger shifts between 
the CO emission peaks, the bandhead minima absorb the peak of the redshifted 
emission components. The sum result of this modeling process is that we
can reproduce the observed CO absorption feature, and that the two 
peaks of the CO feature must be shifted by less than $\Delta \lambda$ = $\pm$
50 \AA. Unfortunately, the large and uncertain velocity broadening does not 
allow us to put useful constraints on the location of the CO emission region.

As shown in Harrison \& Hamilton (2015), carbon is 
highly deficient in the long period CVs SS Cyg, RU Peg, and GK Per (and by 
inference, the numerous other CVs with weak CO features seen in the $K$-band 
spectroscopic surveys of Harrison et al. 2004b, 2005). In contrast, Hamilton et 
al. (2011) found that the secondary stars in several CVs with short orbital 
periods similar to that of WZ Sge appeared to have normal CO absorption 
strengths. They also found that when corresponding UV data exists, CVs with 
weak CO features in the $K$-band had white dwarfs with apparent deficits of
carbon.  Analysis of the minimum light UV spectrum of WZ Sge by Cheng et al. 
(1997) found that the white dwarf appears to have a super-solar abundance of 
carbon. Our results for WZ Sge indicate a donor with normal levels of carbon.

\begin{figure}[htb]
\centerline{{\includegraphics[width=17cm]{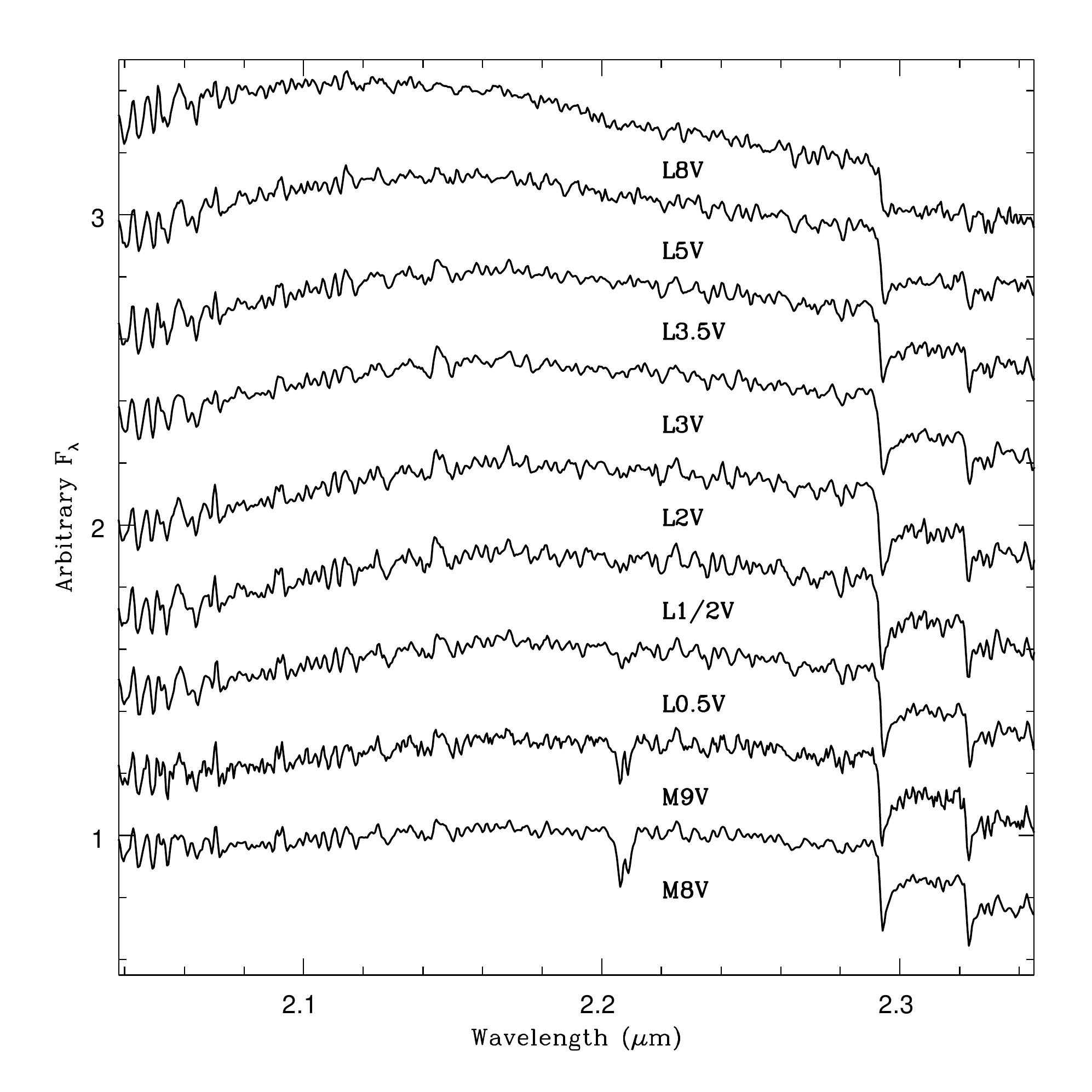}}}
\caption{A $K$-band spectral sequence of objects near, and below, the 
stellar/sub-stellar boundary (data obtained from the IRTF Spectral Library).
For spectral types earlier than L2, the Na I doublet at 2.2 $\mu$m is 
apparent. For the latest L dwarfs, the $^{\rm 12}$CO$_{\rm (3,1)}$ feature
(at 2.322 $\mu$m) becomes very weak.}
\label{irtf}
\end{figure}

Given that we have matched the orbitally averaged spectrum of WZ Sge to an L2 
dwarf, and there is strong evidence for irradiation, the actual spectral type 
of the donor must be later than L2. To attempt to better constrain the 
spectral type of the secondary star we more closely examine the $K$-band 
spectra of L dwarfs. In Fig. \ref{irtf}, we plot spectra 
from the IRTF Spectral Library for spectral types from M8 to L8. The lack of
a Na I doublet (2.2 $\mu$m) at any phase during the orbit confirms that the 
donor must have a spectral type later than L1. The strengths of the CO 
features do not change significantly until the latest L dwarfs. The reason
for this is the small temperature change from one subtype to the next.
As shown in Basri et al. (2000), the temperature decrease from L0 to L4 is
only $\Delta$T = 350 K. By L5, the 
$^{\rm 12}$CO$_{\rm (3,1)}$ bandhead (at 2.322 $\mu$m) has become much weaker, 
and is very weak at L8. It is clear that in the $\phi$ = 0.0 spectrum (Fig.
\ref{ldwarf}), the $^{\rm 12}$CO$_{\rm (3,1)}$ bandhead is quite strong. The 
remainder of the main absorption features in the continua of L dwarfs (or 
opacity minima) show much smaller changes with decreasing temperature. We 
conclude that the spectral type of the secondary in WZ Sge is later than L2, 
but is earlier than L8. Using the temperature scale from Basri et al.
(2000) suggests a temperature for the donor of between 1800 and 2000 K,
consistent with the results of Harrison et al. (2013).

\subsection{The Masses of the Components in WZ Sge}

With the expectation that an ultra-short period CV like WZ Sge is an older 
object, and that its companion was once a more massive star, it is unclear
how closely the donor object properties will match those of a field brown 
dwarf. Since such objects have no internal nuclear energy source, the 
radii of isolated brown dwarfs depend on their age. As discussed in
Sorahana et al. (2013), older brown dwarfs with temperatures near 2000 K
have radii near 0.9 R$_{\rm Jupiter}$, and are expected to have masses of 
0.07 M$_{\sun}$. If we assume this is true for WZ Sge, we can attempt to 
further constrain the parameters of the system. If K$_{\rm 1}$ $\leq$ 37 km 
s$^{\rm -1}$ (Steeghs et al. 2001) and K$_{\rm abs}$ $\geq$ 520 km 
s$^{\rm -1}$, then $q$ $\leq$ 0.071, and M$_{\rm 1}$ $\geq$ 0.98 M$_{\sun}$. 
A more precise value for K$_{\rm 1}$, or a measurement of $v$sin$i$ for the
donor, is desperately needed to further constrain this system.

\section{Discussion and Conclusions}

We have obtained a large set of $K$-band spectra of WZ Sge. When these data
are phased to the orbital period, it becomes obvious that the central dip in 
the CO emission features moved with orbital phase in a pattern that suggested 
absorption by an underlying source. Measuring the position of the 
$^{\rm 12}$CO$_{\rm (2,0)}$ bandhead leads to a radial velocity curve that 
has a semi-amplitude of K$_{\rm abs}$ = 520 $\pm$ 35 km s$^{\rm -1}$. 
This result is consistent with the range of K$_{\rm em}$ values derived by 
Steeghs et al.  (2001) for the motion of the secondary star from 
observations of the irradiation-induced H$\alpha$ emission line. When
the entire data set is phased using our value of K$_{\rm 2}$, the
continuum in the spectrum has features that indicate the donor has a spectral
type near L2. The late spectral type of the donor in WZ Sge, coupled with its
short orbital period, suggest that WZ Sge might be a period bouncer 
(Patterson 1998). Such a classification, however, remains difficult to prove. 
For example, Knigge et al. (2011) suggest that such objects would have very 
low masses: M$_{\rm 2}$ $\leq$ 0.05 M$_{\sun}$. If spectral type correlates 
with mass for these objects, the donors in period-bounce CVs would be expected
to have the spectral characteristics of late-L/early-T brown dwarfs. The L2 
$-$ L5 spectral type we have found for WZ Sge would be too early, and cast 
doubt on its identification as a period bounce system.

While there have been a large number of claims for brown dwarf donors in CVs, 
there is only one secure case for the direct spectroscopic detection of the 
photospheric features of a brown dwarf secondary that we are aware of, and 
that is for the 
secondary in SDSS J1433+1011 (Littlefair et al. 2013). The donor in SDSS 
J1433+1011 also has a spectral type near L2, and is a system that has an 
orbital period very close to the theoretical minimum (P$_{\rm orb}$ = 78.1 
min). The best previous example was the polar J121209.31+013627.7, where low 
resolution infrared spectroscopy is certainly consistent with an L8 donor 
(Farihi et al. 2008). In the other systems with suspected brown dwarf donors, 
it is infrared photometry, eclipses and/or radial velocity studies which 
suggest a sub-stellar companion. Such studies are often dependent 
on the mass and/or temperature of the white dwarf primary, which remain 
challenging to constrain. In the case of WZ Sge, we now have two independent 
radial velocity studies that are consistent with each other and that suggest a 
sub-stellar donor mass, a light curve modeling program that predicted an L2 
donor, and the subsequent direct detection of this object. 

An examination of the near-IR spectra of L dwarfs indicates that the 
$J$-band is a much better wavelength regime for classification than in the 
$K$-band. $J$-band spectroscopy was used by Littlefair et al. (2013) to detect
the L2 donor in SDSS J1433+1011. As shown in McLean et al. (2000), very strong 
absorption features of K I and FeH are present that would allow for a more 
accurate spectral typing of the secondary, and a radial velocity curve could 
be more easily generated from these lines as they would be unhampered by H I 
or molecular emission. WZ Sge has in fact been observed in the $J$-band by 
Littlefair et al. (2000), but the data they presented was not 
Doppler-corrected, and thus any absorption features from the donor would be 
washed-out due to its large motions. Phasing that data set would be useful to 
confirm our results.

\acknowledgements TEH is partially supported by a grant from the NSF 
(AST-1209451). Data presented herein were obtained at the W. M. Keck
Observatory, which is operated as a scientific partnership among the California
Institute of Technology, the University of California, and NASA. The 
Observatory was made possible by the generous financial support of the W. M. 
Keck Foundation.  
\clearpage
\begin{center}
{\bf References}
\end{center}
Basri, G., Mohanty, S., Allard, F., Hauschildt, P. H., et al. 2000, ApJ, 538,
363\\
Cheng, F. H., Sion, E. M., Szkody, P., \& Huang, M. 1997, ApJ, 484, L149\\
363\\
Clarke, F. J., Tinney, C. G., \& Covey, K. R. 2002, MNRAS, 332, 361\\
Cushing, M. C., Rayner, J. T., Vacca, W. D. 2005, ApJ, 623, 1115\\
Dupuy, T. J., Liu, M. C., Bowler, B. P., Cushing, M. C., et al. 2010, ApJ, 721,
1725\\
Farihi, J., Burleigh, M. R., \& Hoard, D. W. 2008, ApJ, 674, 421\\
Hamilton, R. T., Harrison, T. E., Tappert, C., Howell, S. B. 2011, ApJ, 728, 16\\
Harrison, T. E., \& Hamilton, R. T. 2015, AJ, in press (arXiv:1509.03664)\\
Harrison, T. E., Hamilton, R. T., Tappert, C., Hoffman, D. I., et al. 2013,
AJ, 145, 19\\
Harrison, T. E., Osborne, H. L., \& Howell, S. B. 2005, AJ, 129, 2400\\
Harrison, T. E., Johnson, J. J., McArthur, B. E., Benedict, G. F., et al. 2004a,
AJ, 127, 460\\
Harrison, T. E., Osborne, H. L., \& Howell, S. B. 2004b, AJ, 127, 3493\\
Howell, S. B., Harrison, T. E., \& Szkody, P. 2004, ApJ, 602, L49\\
Howell, S. B., Rappaport, S., \& Politano, M. 1997, MNRAS, 287, 929\\
King, A. R., \& Kolb, U. 1995, ApJ, 439, 330\\
Knigge, C., Baraffe, I., \& Patterson, J. 2011, ApJS, 194, 28\\
Kolb, U., \& Baraffe, I. 1999, MNRAS, 309, 1034\\
Littlefair, S. P., Savoury, C. D. J., Dhillon, V. S., Marsh, T. R., et al. 
2013, MNRAS, 431, 2820\\
Littlefair, S. P., Dhillon, V. S., Howell, S. B., \& Ciardi, D. R. 2000, MNRAS,
313, 117\\
Liu, M. C., \& Leggett, S. K. 2005, ApJ, 634, 616\\
Maiolino, R., Rieke, G. H., \& Rieke, M. J. 1996, AJ, 111, 537\\
McLean, I. S., Mavourneen, K. W., Becklin, E. E., Figer, D. F., et al. 2000,
ApJ, 544, L45\\
Osaki, Y. 1996, PASP, 108, 39\\
Paczy\'{n}ski B., 1981, Acta Astr., 31, 1\\
Patterson, J., Gianluca, M., Richmond, M. W., Martin, B., et al. 2002, PASP,
114, 721\\
Patterson, J., Richman, H., \& Kemp, J. 1998, PASP, 110, 403\\
Patterson, J. 1998, PASP, 110, 1132\\
Skidmore, W., Wynn, G. A., Leach, R., \& Jameson, R. F. 2002, MNRAS, 336, 1223\\
Sorahana, S., Yamamura, I., \& Murakami, H. 2013, ApJ, 767, 77\\
Spruit, H. \& Rutten, R. 1998, MNRAS, 299, 768\\
Steeghs, D., Marsh, T., Knigge, C., Maxted, P. F. L., et al. 2001, ApJ, 562, L145\\
Thorstensen, J. R., 2003, AJ, 126, 3017\\

\clearpage
\clearpage
\clearpage
\clearpage
\clearpage
\clearpage
\clearpage

\end{document}